\documentclass{emulateapj}

\newcommand{\ha}{\mbox{H$\alpha$}}

\def\h2{H$_2$}

\def\c2{cm$^{-2}$}
\def\2e{$^{2}$\ }
\def\2{$^{2}$}
\def\3e{$^{3}$\ }
\def\3{$^{3}$}
\def\-1e{$^{-1}$\ }
\def\-1{$^{-1}$}
\def\-2e{$^{-2}$\ }
\def\-2{$^{-2}$}
\def\-3e{$^{-3}$\ }
\def\-3{$^{-3}$}

\shorttitle{The H$_2$-Pressure Relation}
\shortauthors{Blitz \& Rosolowsky}
\begin{document}

\title{The Role of Pressure in GMC Formation II: The H$_2$ - Pressure
Relation}
\author{Leo Blitz and Erik Rosolowsky\altaffilmark{1,2}}
\affil{Department of Astronomy, University of California at Berkeley,
\\601 Campbell Hall MC 3411, CA 94720}
\email{blitz@astro.berkeley.edu}
\keywords{Galaxies:ISM --- ISM:evolution --- ISM:molecules}
\altaffiltext{1}{Current Address: Center for Astrophyiscs, 60 Garden St. MS-66, Cambridge, MA 02138}
\altaffiltext{2}{NSF Astronomy \& Astrophysics Postdoctoral Fellow}
\begin{abstract}
We show that the ratio of molecular to atomic gas in galaxies is
determined by hydrostatic pressure and that the relation between the
two is nearly linear.  The pressure relation is shown to be good over
three orders of magnitude for 14 galaxies including dwarfs,
\ion{H}{1}-rich, and \h2-rich galaxies as well as the Milky Way.  The
sample spans a factor of five in mean metallicity.  The rms scatter of
individual points of the relation is only about a factor of two for
all the galaxies, though some show much more scatter than others.
Using these results, we propose a modified star formation
prescription based on pressure determining the degree to which the ISM
is molecular. The formulation is different in high and low pressure
regimes defined by whether the gas is primarily atomic or primarily
molecular. This formulation can be implemented in simulations and
provides a more appropriate treatment of the outer regions of spiral
galaxies and molecule-poor systems such as dwarf irregulars and
damped Ly$\alpha$ systems.
\end{abstract}

\section{Introduction} Millimeter-wave and infrared observations have
long established that all stars form in molecular clouds; most of
these form in Giant Molecular Clouds (GMCs).  This conclusion has good
theoretical support because Jeans instabilities with masses in the
range of 0.1 -- 100 $M_\sun$ occur in regions with temperatures and
densities found only in molecular clouds.  Although good progress has
been made in understanding how low-mass stars from, the physics of
high-mass star and cluster formation remains elusive.  Yet, nearly all
of the information we have about present-day star formation in other
galaxies comes from high mass stars.  As a result, it would seem that
understanding star formation in other galaxies would be a daunting
task.

Despite our ignorance, it may not be necessary to know the details of
star formation to understand how stars form on galactic scales.  For
example, a number of studies suggest that in the disks of normal
galaxies, the efficiency of star formation --- the fraction of the
molecular mass of a galaxy turned into massive stars within $10^8$ yr
--- shows an rms variation of only a factor of 2 when averaged over
kiloparsec scales \citep[e.g.~][etc.]{mur02}.  This constancy suggests
that understanding star formation on kiloparsec or larger scales
within galaxies may reduce largely to a question of how molecular gas
and thus how GMCs form. That is, once GMCs form, the average rate of
star formation is determined to within a factor of 2 as long as the
IMF is relatively constant. Thus, finding a prescription for how GMCs
form in a galaxy, together with a robust value of the star formation
efficiency, may be all that is needed to determine the global star
formation history of a galaxy. Work of this sort has been pioneered by
Kennicutt (1983, 1989, 1998), but his inclusion of atomic gas, which
is inert to star formation, is unsatisfactory, since no known stars
form from atomic gas.  Recently, however, \citet{krumholz05}, have
provided a treatment that justifies including atomic gas in the
\citet{k98} star formation prescription.

Following  work by \citet{wb02}, \citet[][hereafter
BR04]{pressure1} investigated whether the mean ratio of molecular to
atomic hydrogen, $R_{mol}$, at a given radius in a spiral galaxy is
determined primarily by a single parameter: the interstellar gas
pressure, $P_{ext}$. They showed that this hypothesis leads to a
prediction that the transition radius, $R_{t}$, where $R_{mol}$ = 1,
should occur at a constant value of $\Sigma_\star$, the {\it stellar}
surface density. For a sample of 30 galaxies, the stellar surface
density at the transition radius ($\Sigma_{\star,t}$) is constant to
within 40\%, consistent with the hypothesis that $R_{mol}$ is a
function ($f$) of $P_{ext}$ alone, i.e. $R_{mol}=f(P_{ext})$.  In this
paper we extend the work of \citet{wb02} with a pixel-by-pixel
analysis using additional galaxies to find the functional form of
$f(P_{ext})$ over a much larger range of pressure.  We show that rms
deviations from the derived relation are no more than about a factor
of 2 for a range of galaxy types and metallicities.~  Finally, we use
the pressure-molecular fraction relation to derive a star-formation
law that modifies the \citet{k98} star formation prescription,
particularly at low molecular gas fractions.

\section{Background}

\label{background}

BR04 estimate $P_{ext}$ for disk galaxies from the midplane pressure
in an infinite, two-fluid disk with locally isothermal stellar and gas
layers.  They assume that the gas scale height is much less than the
stellar scale height, which is typical of disk galaxies; so that, to
first order:
\begin{equation}
P_{ext} = (2G)^{0.5}\Sigma_g v_g\left[{\rho_\star}^{0.5} + \left(\frac
{\pi}{4}
\rho_g\right)^{0.5}\right].
\label{fullpressure}
\end{equation}

\noindent Here, $\Sigma_{g}$ is the total surface density of the gas,
$v_g$ is the vertical velocity dispersion of the gas, $\rho_\star$ is
the midplane volume density of stars, and $\rho_g$ is the midplane
volume density of gas.

In most galaxy disks, $\rho_\star$ is much larger than $\rho_g$ when
azimuthally averaged.  For a self-gravitating stellar disk, the
stellar surface density, $\Sigma_\star = 2\rho_\star h_\star$, where
$h_\star$ is the stellar scale height and $h_\star = ({v_\star}^2/2\pi
G
\rho_\star)^{0.5}$.  Thus, neglecting $\rho_g$,
Equation 1 becomes:

\begin{equation}
P_{ext} =  0.84 (G \Sigma_\star)^{0.5}\Sigma_g \frac {v_g} {(h_\star)^
{0.5}}. \
\label{approxpressure}
\end{equation}

\noindent   Direct solution of the fluid equations by numerical
integration shows that this approximation is accurate to 10\% for
$\Sigma_\star > 20~ M_{\odot}\mbox{ pc}^{-2}$ (where $\rho_\star
\lesssim \rho_g$), which covers the range of stellar surface  
densities in this
study.

We choose to express the midplane pressure in the form of Equation
\ref{approxpressure} since there is good evidence that both $h_\star$
and $v_g$ are constant with radius in disk galaxies (See BR04 for
references).  Furthermore, because of the weak dependence of $P_{ext}$
on $h_\star$, and the small variation of $h_\star$ measured from
galaxy-to-galaxy \citep{kregel-stellar}, we expect variations of $h_
\star$ to have little effect on $P_{ext}$.

By assumption,
\begin{equation}
R_{mol}\equiv \Sigma(\mbox{H}_2)/\Sigma(\mbox{\ion{H}{1}}) = f(P_{ext})
\label{gequation}
\end{equation}

\noindent Thus, since $(v_g/\sqrt {h_\star})$ is approximately constant
within galaxies, then

\begin{equation}
\Sigma(\mbox{H}_2)/\Sigma(\mbox{\ion{H}{1}}) =
f\left[P_{ext}(\Sigma_g,\Sigma_\star)\right]
\end{equation}

We wish to find $f$ from Eq. \ref{approxpressure} for as many galaxies
and galaxy types as possible.

\section{Analysis}
\label{analysis}

\subsection{Data}
We have collected a sample of 14 galaxies for which the atomic,
molecular and stellar surface densities can be determined.  The
galaxies and their adopted orientation
parameters are listed in Table
\ref{galprops}.  We also list the original publication of the
\ion{H}{1} and CO maps.  Most of the \ion{H}{1} data are from the VLA
and the most of the CO data are from the BIMA Survey of Nearby
Galaxies \citep{song2}.  For all of the galaxies in Table
\ref{galprops} except IC10 and the Milky Way,
we derive their stellar surface densities from the 2~$\mu$m
($K_s$-band) image in the 2MASS Large Galaxy Atlas \citep{2mass-lga}.
We augment this sample with the 2MASS $K_s$ image used in
\citet{leroy-ic10} for IC10 and data from the literature for the
Milky Way.

\begin{deluxetable*}{cccccccc}
\tablecaption{\label{galprops}Galactic Parameters}
\tablewidth{0pt}
\tablehead{
%\begin{tabular}{cll}
\colhead{Galaxy Name} & \colhead{Distance} &\colhead{Inc.} & \colhead
{P.A.} &
\colhead{{\sc H~i} data} & \colhead{CO data} & \colhead{Resolution} &
\colhead{Stellar Scale}\\
     & & & & & & & \colhead{Height}\\
\colhead{}& \colhead{(Mpc)}& \colhead{($^{\circ}$)}& \colhead{($^
{\circ}$)} &
\colhead{reference} & \colhead{reference} & \colhead{(kpc)} & \colhead
{(kpc)}\\
}
\startdata
%IC  342  & 3.9 &  31 & 37  & 1  & 2  & 720  \\
NGC 598	 & 0.85 & 56 & 203 & 1	& 2  & 0.26&   0.21\\
NGC 3521 & 7.2 &  58 & 164 & 3  & 4  & 1.3 &   0.23\\
NGC 3627 & 11.1 & 63 & 176 & 5  & 4  & 1.3 &   0.39\\
NGC 4321 & 16.1 & 32 & 154 & 6	& 4  & 1.6 &   0.50\\
NGC 4414 & 19.1	& 55 & 159 & 7	& 4  & 1.6 &   0.31\\
NGC 4501 & 16.0 & 63 & 140 & 6	& 8  & 1.8 &   0.41\\
NGC 4736 & 4.3 &  35 & 100 & 9	& 4  & 0.31 &  0.11\\
NGC 5033 & 18.6 & 62 & 170 & 10	& 4  & 2.1 &   0.25\\
NGC 5055 & 7.2 &  55 & 105 & 11	& 4  & 0.45  & 0.24\\
NGC 5194 & 7.7 &  15 &  0 & 12	& 4  & 0.30  & 0.33\\
NGC 5457 & 7.4 &  27 & 40 &  11  & 4  & 0.20 & 0.39\\
NGC 7331 & 15.1 & 62 & 172 & 13	& 4  & 0.45  & 0.28\\
IC 10 & 0.95 & 48 & 132 & 14 & 15 & 0.28 & 0.30\\
\enddata
\tablerefs{(1) \citet{deul}; (2) \citet{m33-fcrao};
(3) M. Thornley, private
communication; (4) \citet{song2};  (5) \citet{alexander-3627}; (6)
\citet{knapen-4321}; (7) \citet{thornley-4414}; (8) \citet{wbb04}; (9)
\citet{braun-superclouds}; (10) \citet{thean-5033}; (11)
\citet{thornley-5055}; (12) \citet{m51-h1}; (13) \citet{spitzer-7331};
(14) \citet{wil-ic10} (15) \citet{leroy-ic10}}
\end{deluxetable*}

\subsection{Derivation of Physical Quantities}
\label{derivations}
We measure the variation of molecular gas fraction ($R_{mol}\equiv
\Sigma_{\mathrm{H2}}/\Sigma_{\mathrm{HI}}$) as a function of midplane
hydrostatic pressure on a pixel-by-pixel basis from 2-dimensional maps
of these quantities for the 14 galaxies in Table \ref{galprops}.  We
estimate the midplane pressure in each galactic disk using an
alternative form of the expression of BR04:
\begin{eqnarray}
\nonumber
\frac{P_{ext}}{k} &=& 272 \mbox{ cm$^{-3}$ K}
\left(\frac{\Sigma_{g}}{M_{\odot}\mbox{ pc}^{-2}}\right)
\left(\frac{\Sigma_{\star}}{M_{\odot}\mbox{ pc}^{-2}}\right)^{0.5}\\
&\times&\left(\frac{v_g}{\mbox{km s}^{-1}}\right)
     \left(\frac{h_\star}{\mbox{pc}}\right)^{-0.5}.
\label{pext}
\end{eqnarray}
%This formulation is appropriate
%for the disks of galaxies where the stellar surface density is at
%least a factor of two larger than the neutral gas surface density.

We determine the surface densities of the atomic gas, the molecular
gas, and the stars from the maps of 21-cm, CO$(J=1\to 0)$, and
$K_s$-band emission respectively.  First, we remove foreground stars
from the $K_s$ image by isolating point sources and replacing them
with the median sky brightness surrounding each source.  We then
convolve the maps to the coarsest resolution among the three images
(usually the
\ion{H}{1}).  The resulting resolution measured in the plane of the
sky is given in Table \ref{galprops}.  We then flag all pixels in the
resulting maps that have less than a $5\sigma_{rms}$ significance in
any of the integrated-intensity maps and do not consider these points
in our analysis.  Finally, we resampled the images onto the same
coordinate grid so that the pixels in different images can be compared
directly.

Using the resulting maps, we estimate the surface densities of the
various galactic components.  For the atomic gas, the surface density
is:
\begin{equation}
\label{h1eq}
\left(\frac{\Sigma_{\mathrm{HI}}}{M_{\odot}\mbox{ pc}^{-2}}\right)
=0.020\left(\frac{I_{\mathrm{HI}}}{\mbox{K km s}^{-1}}\right)\cos i,
\end{equation}
where $i$ is the inclination of the galactic disk.  We have assumed a
mean particle mass of 1.36~$m_{H}$ to account for the presence of
helium and metals.  Similarly, the surface density of molecular gas is
\begin{equation}
\label{h2eq}
\left(\frac{\Sigma_{\mathrm{H2}}}{M_{\odot}\mbox{ pc}^{-2}}\right)
=4.4\left(\frac{I_{\mathrm{CO}}}{\mbox{K km
s}^{-1}}\right)\cos i,
\end{equation}
adopting a CO-to-H$_2$ conversion factor of
$N(\mbox{H}_2)/I_{\mathrm{CO}} = 2\times 10^{20} \mbox{ cm}^{-2}
(\mbox{K} \mbox{ km s}^{-1})^{-1}$ and a mean particle mass of
1.36~$m_{H}$ per hydrogen nucleus.  Finally, we determine the stellar
surface density from integrated $K_s$-band. light.  We adopt a stellar
mass-to-light ratio of $M_{K}/L_{K} = 0.5 M_{\odot}/L_{\odot}$ which
approximates a large variety of star formation histories to 20\%
accuracy \citep{m2l-bell}.  Then
\begin{equation}
\label{starseq}
\log_{10}\left(\frac{\Sigma_{\star}}{M_{\odot}\mbox{ pc}^{-2}}\right) =
-0.4\mu_{Ks}+9.62+\log_{10}\cos i,
\end{equation}
where $\mu_{Ks}$ is the surface brightness of the galaxy in magnitudes
per square arcsecond.

Calculating $P_{ext}$ requires knowledge of the gas velocity
dispersion and the stellar scale height.  For all galaxies, we assume
a constant gas velocity dispersion of $v_g=8$~km~s$^{-1}$, which is
observed across the disks of several galaxies
\citep[e.g.][]{svdk,dhh90,mw-h1,malhotra-h1}.  We estimate the stellar
scale height $h_\star$ based on the relationship between the radial
scale length of the stellar disks ($R_\star$) and the corresponding
stellar scale heights found by \citet{kregel-stellar}.  Fitting their
data gives
\begin{equation}
\label{scaleht}
\log_{10}h_\star = (-0.23\pm 0.05)+(0.8\pm 0.1)\log_{10} R_\star.
\end{equation}
where $h_\star$ and $R_\star$ are measured in parsecs.  To measure
$R_\star$, we azimuthally average the stellar surface density in
annuli of constant galactocentric radius assuming the orientation
parameters in Table \ref{galprops}.  We fit an exponential function to
the resulting profile ignoring regions indicating the presence of
a bulge.  Using the derived value of $R_\star$, we then calculate the
scale height using equation \ref{scaleht}.  The derived stellar scale
heights
are reported in Table \ref{galprops}.  The estimate of the scale
height is calculated prior to convolution to a common resolution.
We assume that that the ionized
gas contributes negligibly to $\Sigma_g$; in any event, its scale
height is large \citep{reynolds-scaleht}, and its contribution to the
midplane pressure should be small.

\begin{figure}
\begin{center}
\plotone{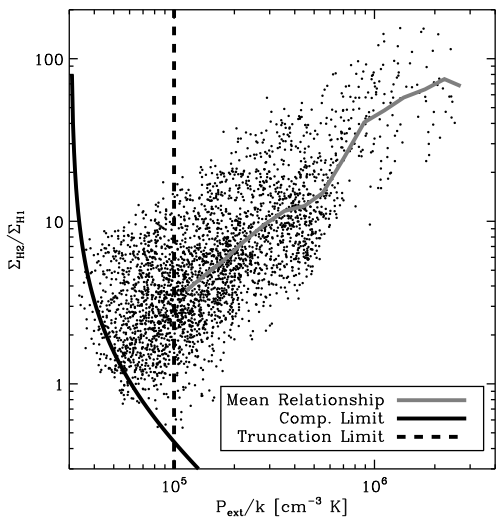}
\caption{\label{m51fig} Molecular gas fraction as a function of
midplane hydrostatic pressure for NGC 5194 (M51).  The mean
relationship is
plotted in a thick gray line.  The completeness limit arising from the
molecular gas observations is plotted as a solid black line.  The
intersection of this line with the bottom of the observed distribution
determines the truncation limit.  The relationship between $P_{ext}$
and $R_{mol}$ is only calculated above this limit.}
\end{center}
\epsscale{1.0}
\end{figure}

\subsection{Completeness}

We calculate $R_{mol}$ as a function of pressure on a pixel-by-pixel
basis for all pixels that have measured surface densities of
\ion{H}{1}, CO, and 2~$\mu$m emission.  Figure \ref{m51fig} is a plot
of the correlation between these two quantities for NGC 5194
(M51). The two quantities are clearly correlated, but the different
signal-to-noise ratios of the three data sets affect the distribution
of points in the Figure.  Of the three tracers, the CO emission has
the lowest signal-to-noise, and is found at the fewest positions.  It
is thus the limiting factor in our analysis.

\begin{figure*}
\begin{center}
\epsscale{1.0}
\plotone{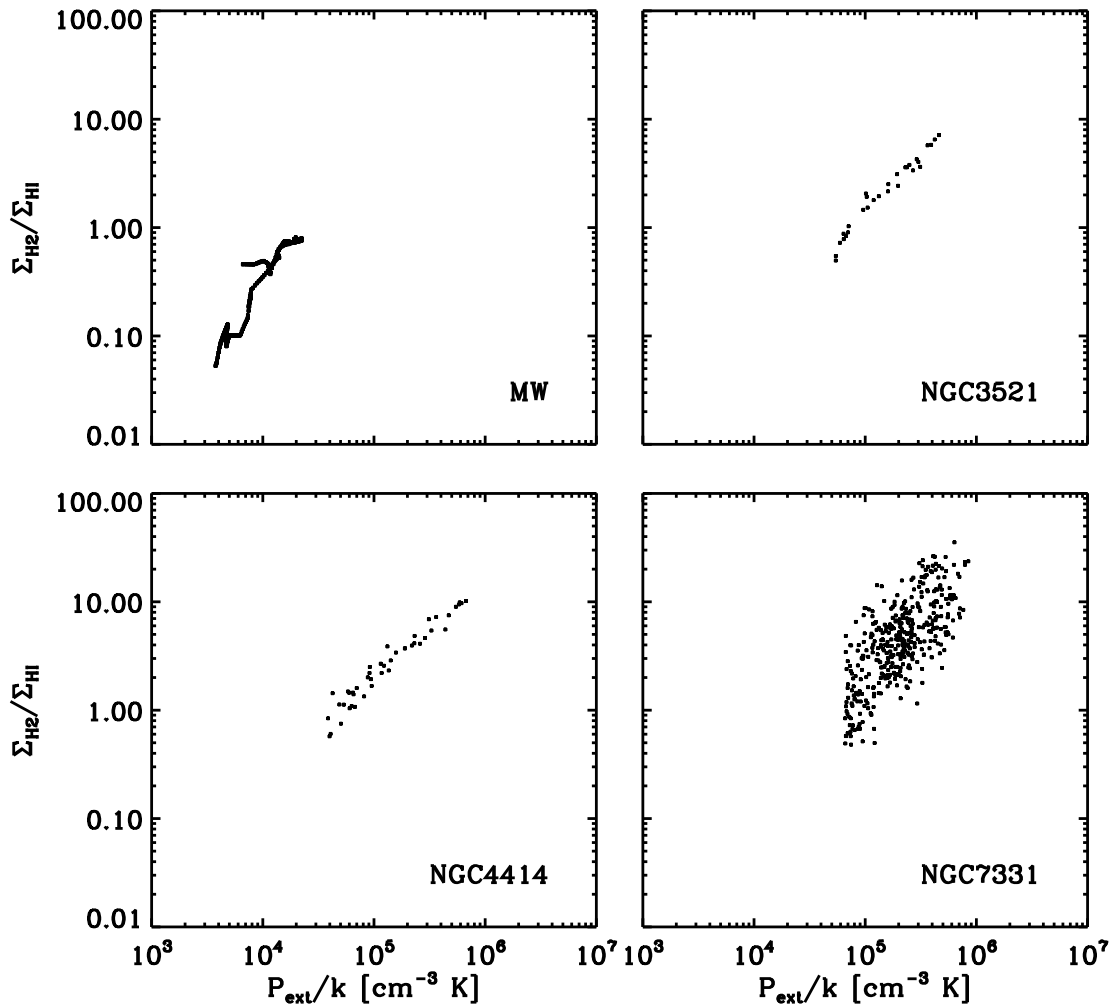}
\caption{\label{examples} Plots of $P_{ext}$ vs. $R_{mol}$ for four
galaxies showing a range of scatter on a pixel-by-pixel basis. For
galaxies such as NGC 3521, the scatter is so small (15\%) that finding
the completeness limit is trivial.  For galaxies such as NGC 7331, we
must use the technique described in the text for estimating the
completeness limits in the relationship.}
\end{center}
\epsscale{1.0}
\end{figure*}

To illustrate the effects of finite sensitivity, we plot the locus of
points where $\Sigma_{\mathrm{H2}}=8~M_{\odot}~\mbox{pc}^{-2}$ (the
5$\sigma$ sensitivity limit),
$\Sigma_\star=300~M_{\odot}~\mbox{pc}^{-2}$, and a range of
$\Sigma_{\mathrm{HI}}$ as the solid curve in Figure \ref{m51fig}. We
choose $\Sigma_\star=300~M_{\odot}~\mbox{pc}^{-2}$, so that the curve
roughly coincides with the left boundary of the data. The curve is a
good
match to the shape of the left edge of the data in Figure
\ref{m51fig}.  At a given pressure, points that lie below the line
have CO surface brightnesses that are too weak to be detected; the
solid line therefore represents the completeness limit for the
observations.  Because we wish to bin the data in values of $P_{ext}$,
we have complete CO data only when the lowest measured values of
$R_{mol}$ for the entire sample lie above the completeness line.  Thus
all of the data to the right of the dotted line, the intersection of
the completeness line with the lowest values of $R_{mol}$ have
complete CO data (i.e. complete measured values of
$\Sigma_{\mathrm{H2}}$).  Plots of $P_{ext}$ vs. $R_{mol}$ for four
other representative galaxies are shown in Figure \ref{examples},
showing that the scatter of the individual points within a galaxy
exhibits considerable variation.

\subsection{The Relationship Between $P_{ext}$ and $R_{mol}$}
\label{fitsec}
Since we wish to measure the molecular gas fraction as a function of
$P_{ext}/k$, we average $R_{mol}$ in bins of 0.1 dex in $P_{ext}/k$.
To avoid biases due to the finite sensitivity, the analysis is
restricted to $P_{ext}/k$ larger than $P_{min}/k$, the value where the
completeness curve intersects the bottom of the data distribution.
Practically, we define $P_{min}/k$ as the bin containing the minimum
value of $R_{mol}$ in the sample, and only consider bins of
$P_{ext}/k$ with this value and larger.  Since the upper limits on the
CO surface brightness affect both $P_{ext}$ and $R_{mol}$, points
without identifiable CO emission provide little additional
information, and we discard these data from our analysis. The
uncertainty in $\log R_{mol}$ ($\delta \log R_{mol}$) is determined by
the standard deviation of points in each bin ($\sigma_{\log R_{mol}}$)
divided by the square root of the number of independent samples in
each bin:
\begin{equation}
\delta \log R_{mol} \equiv \frac{\sigma_{\log R_{mol}}}{\sqrt{N/n_
{beam}}},
\end{equation}
where $N$ is the number of pixels contributing to each bin and
$n_{beam}$ is the number of pixels per independent beam.

\begin{figure*}
\begin{center}
\plotone{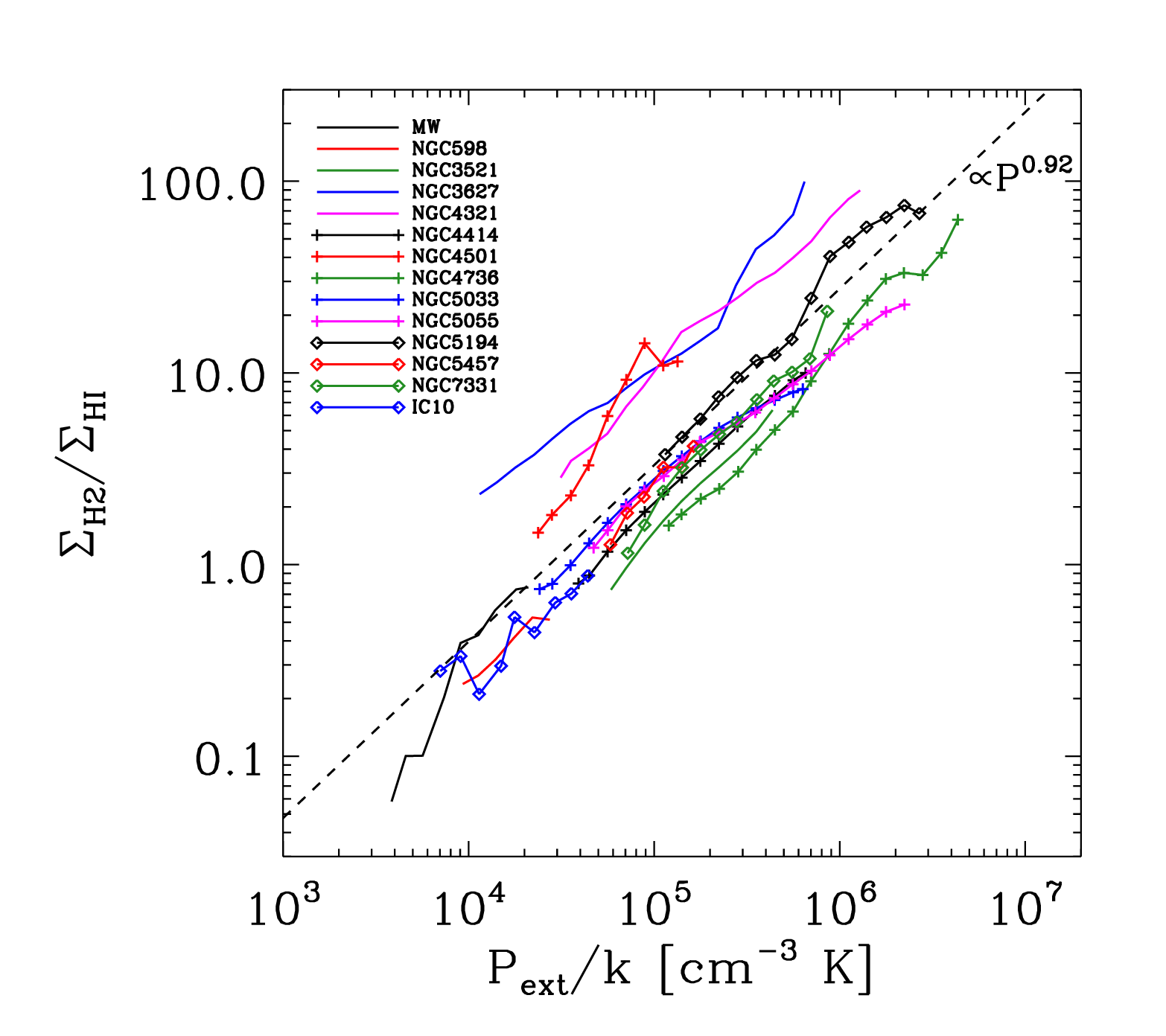}
\caption{\label{pressplot} Molecular gas fraction as a function of
midplane hydrostatic pressure for 14 galaxies.  All galaxies follow a
similar scaling between $P_{ext}$ and $R_{mol}$.  The mean index of
the relationship is $0.92\pm 0.07$ (dashed line).   There are
two categories of galaxies, offset from each other by a factor
of 5 in $P_0$.}
\end{center}
\end{figure*}

\begin{deluxetable*}{cccccccc}
\tablecaption{\label{results}Results of Pressure Analysis}
\tablewidth{0pt}
\tablehead{
%\begin{tabular}{cll}
\colhead{Galaxy} & \colhead{$\alpha$} & \colhead{$P_0/k$} &
\colhead{Scatter\tablenotemark{a}} &
\colhead{$\langle \Sigma_{\mathrm{HI}}\cos i\rangle$} &
\colhead{Morphological}\\
\colhead{Name}&  \colhead{} & \colhead{($10^4$ cm$^{-3}$ K)} & &
\colhead{($M_{\odot}~\mbox{pc}^{-2}$)} & \colhead{class\tablenotemark
{b}}
}
\startdata
MW & $1.64\pm 0.11$ & 2.0 & 0.09 & 8.4 & Sb \\
NGC 598 & $0.87\pm 0.14$ & 5.1 & 0.03 & 9.3 & SA(s)cd\\
NGC 3521 & $1.02\pm 0.03$ & 7.1 & 0.02 & 16.8 & SAB(rs)bc\\
NGC 3627 & $0.81\pm 0.03$ & 0.4 & 0.10 & 4.3 &  SAB(s)b\\
NGC 4321 & $0.84\pm 0.03$ & 0.7 & 0.06 & 6.7 & SAB(s)bc\\
NGC 4414 & $0.89\pm 0.02$ & 4.6 & 0.02 & 14.4 & SA(rs)c?\\
NGC 4501 & $1.07\pm 0.08$ & 1.2 & 0.13 & 4.2 & SA(rs)b\\
NGC 4736 & $0.93\pm 0.04$ & 6.5 & 0.09 & 11.1 & (R)SA(r)ab\\
NGC 5033 & $0.76\pm 0.03$ & 3.0 & 0.05 & 11.5 & SA(s)c\\
NGC 5055 & $0.73\pm 0.02$ & 2.8 & 0.03 & 11.5 & SA(rs)bc\\
NGC 5194 & $1.00\pm 0.05$ & 3.0 & 0.07 & 13.2 & SA(s)bc pec\\
NGC 5457 & $0.58\pm 0.32$ & 2.1 & 0.09 & 16.6 & SAB(rs)cd\\
NGC 7331 & $1.01\pm 0.06$ & 5.1 & 0.05 & 14.5 & SA(s)b\\
IC 10 & $0.73\pm 0.14$ & 5.6 & 0.10 & 6.4 & dIrr IV/BCD\\
\hline
Mean & $0.92\pm 0.07$ & 3.5 & 0.06 & 10.6 & Sbc\\
Mean (Non-interacting)\tablenotemark{c} &  $0.92\pm 0.10$ & 4.3 & 0.05 &
12.2 & Sbc\\
\hline
Combined Data & $0.94\pm 0.02$ & 4.5 & 0.14 & 9.9& \nodata \\
\enddata
\tablenotetext{a}{Defined as the standard deviation of the residuals:
$\Delta \log R_{mol} \equiv \log R_{mol}-\alpha(\log P_{ext} - \log
P_0)$}
\tablenotetext{b}{From RC3 \citep{rc3}.}
\tablenotetext{c}{Excludes data from NGC 3627, NGC 4321 and NGC 4501.}
\end{deluxetable*}

We plot the relationship between hydrostatic pressure and molecular
gas fraction in Figure \ref{pressplot}. The figure shows that a power
law is a good representation of the functional form between the two
quantities.  The plots for the individual galaxies have a similar
slope but seem to lie in two groups. We fit a linear relationship
between $\log (P_{ext}/k)$ and $\log R_{mol}$ to estimate the
parameters for a power-law relationship between the two variables:
\begin{equation}
R_{mol} = \left(\frac{P_{ext}}{P_0}\right)^\alpha
\label{generic-fcn}
\end{equation}
The derived values of the index $\alpha$ are reported in Table
\ref{results} with uncertainties generated by bootstrapping. $P_0$ is
the external pressure in the ISM when the molecular fraction is unity.
The results for all of the galaxies is given by:
\begin{equation}
\label{allgals}
R_{mol} = \left[\frac{P_{ext}/k}{(3.5\pm 0.6)\times 10^4}
\right]^{0.92\pm 0.07}.
\end{equation}

The data for NGC 3627, NGC 4321, and NGC 4501 lie significantly above
the data for the remainder of the sample.  Splitting the data into two
groups, we find $\langle P_0/k\rangle = 7700 \mbox{ cm}^{-3}~\mbox{K}$
for these three galaxies but $\langle P_0/k\rangle = 43000 \mbox{
cm}^{-3}~\mbox{K}$ for the remainder of the sample.  However, the mean
index, $\langle\alpha \rangle$, for each of the subsets is
indistinguishable from that of the sample as a whole.  The difference
between the two populations can be attributed to differences in the
\ion{H}{1} content of the disks.  In Table \ref{results}, we also
tabulate the median values of $\Sigma_{\mathrm{HI}}\cos i$ used in the
analysis.  We note that $\langle \Sigma_{\mathrm{HI}}\cos i\rangle$ is
significantly lower for these three galaxies than for the remainder of
the systems.  The low values of $\Sigma_{\mathrm{HI}}\cos i$ likely
arise from ram pressure or tidal stripping in these galaxies.  The
other galaxies in our sample appear to be field galaxies without
significant tidal or ram pressure interactions\footnote{NGC 5194 (M51a)
is interacting with NGC 5195 (M51b), but the interaction shows no  
evidence of
stripping neutral gas from NGC 5194.}.  NGC 3627 is in the Leo Triplet
of galaxies and has had a recent interaction with NGC 3628, evidenced
by the asymmetric
\ion{H}{1} distribution in NGC 3627 and the long tail of \ion{H}{1}
from NGC 3628 \citep{n3627-h1}.  NGC 4321 and NGC 4501 are both
members of the Virgo Cluster and are known to be \ion{H}{1} poor from
previous work \citep{virgo-h1}.  The \ion{H}{1} morphology of NGC 4501
shows ample evidence that the galaxy is undergoing ram pressure
stripping in the high-pressure cluster environment.  NGC 4321 is less
obviously disturbed, but still \ion{H}{1} poor.  The three galaxies do
not appear to be discrepant in any other way.

If we remove these galaxies from the mean relationship, we obtain:

\begin{equation}
\label{finalpress}
R_{mol} = \left[\frac{P_{ext}/k}{(4.3\pm 0.6)\times 10^4}
\right]^{0.92\pm 0.10},
\end{equation}

\noindent
which changes $P_0$ by only 20\%.  This may be a more representative
relationship than Equation {\ref{allgals} because the discrepant
galaxies may be subject to pressure in addition to the hydrostatic
pressure, which would move them to the right in Figure
\ref{pressplot}.  We also fit a power law to the combined data for the
non-interacting galaxies, rather than fitting each galaxy and then
averaging (Figure \ref{alldata}).  This weights the fits by the total
number of data at a given $P_{ext}$.  Because the nearby,
\ion{H}{1}-rich galaxies (IC10, M33) have many more independent
samples, this fit emphasizes galaxies with low values of $R_{mol}$.
Nonetheless, the resulting fit (Combined Data in Table \ref{results})
is indistinguishable from the average of the individual fits to the
non-interacting data.

It is worth noting that although we derive a relationship
involving the gas pressure, because we take the velocity dispersion of
the gas as constant, and because the turbulent pressure, $\rho v^2$,
provides most of the support of the gas layer, our expression is
essentially equivalent to $R_{mol}$ being a function of the midplane
gas density.  We cannot, with our data alone, decide between these two
possibilities.  But using the equivalent density parameterization
yields a true Schmidt law, where the star formation rate is dependent on
ISM density \citep{schmidt59}.

\begin{figure}
\begin{center}
\plotone{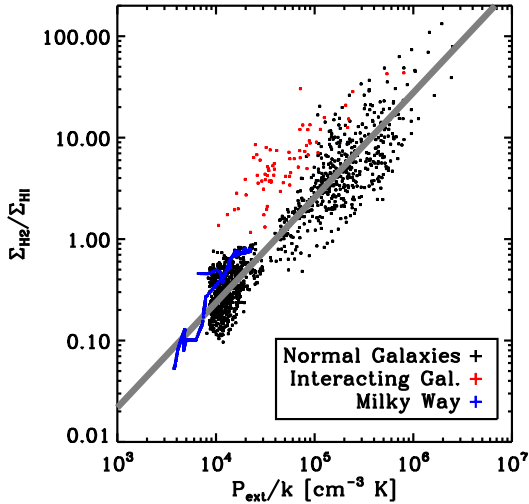}
\caption{\label{alldata} Molecular gas fraction as a function of
midplane hydrostatic pressure for all data in the sample.  Each point
represents an independent measurement of $\Sigma_{\mathrm{HI}}$,
$\Sigma_{\mathrm{H2}}$ and $\Sigma_{\star}$.  A linear
fit to $\log R_{mol}$ vs. $\log P_{ext}/k$ is shown as a thick gray
line.  The data for the interacting galaxies (red, NGC 3627, NGC 4321,
NGC 4501) are excluded from the fit.  The results are
indistinguishable from determining the relationship between $R_{mol}$
and $P_{ext}$ for each galaxy individually.}
\end{center}
\end{figure}

\subsection{Comparison with Previous Work}

\citet{wb02} found, for a sample of seven galaxies, that $R_{mol}$
$\propto {P_{ext}}^{0.8}$ over one and a half orders of magnitude in
pressure. For a given $R_{mol}$, $P_{ext}$ ranges over an order of
magnitude suggesting that the dispersion in the pressure is about a
factor of 2 -- 3.  We have made several improvements to the \citet{wb02}
analysis as follows.  Rather than taking azimuthal averages, we have
calculated $R_{mol}$ and $P_{ext}$ on a pixel-by-pixel basis.
\citet{wb02} used an expression for the pressure by \citet{eg93- 
molfrac} that
requires knowledge of the stellar velocity dispersion in disks, a
quantity that is difficult to measure, and one that is expected to
vary significantly as a function of radius. In this paper, we use a
different, but equivalent expression that relates the pressure to the
stellar surface density and stellar scale height (BR04).  The surface
density is observable; the scale height varies little within
galaxies and its value can be inferred from the observable scale
length \citep[Equation \ref{scaleht}]{kregel-stellar}. By including
additional galaxies, we are able to extend \citet{wb02} to a total of
three orders of magnitude and to include lower mass and lower
metallicity galaxies. Our method also significantly reduces the
scatter among the galaxies, measured by the dispersion of the values
of $P_0 /k$ in Table \ref{results}.

We note that over the range of galactic radius for M33 included
in our sample, there is a decrease in the metallicity by a factor of 6
\citep{hh95}.  Although the range of pressure sampled in M33 is small,
there is no systematic offset from the mean relation shown in Figure
\ref{pressplot} with metallicity.  The same is true for IC10, which
has a metallicity 1/5 solar \citep{garnett90}.

On theoretical grounds, \citet{eg93-molfrac} derives a much steeper
dependence of $R_{mol}$ on pressure than we do, but he also includes a
dependence on the emissivity of the dissociating radiation field, $j$.
\citet{wb02} found that if one assumes that $j \propto \Sigma_{gas}
\propto \Sigma_\star$, Elmegreen's (1993) expression is equivalent to
$f_{mol}= {\Sigma_{\mathrm{H2}}}/{{\Sigma_{g}}} \propto {P_{ext}}^ 
{1.7}$.
However, \citet{wb02} did not take
into account that $c_\star$, the velocity dispersion of the stars, has a
radial dependence on $\Sigma_\star$, which is approximately $\propto
{\Sigma_\star}^{0.5}$.  By ignoring the gravity of the gas as we have  
done
in equation \ref{approxpressure}, and including the dependence of
$c_\star$ on ${\Sigma_\star}, $ Elmegreen's expression for
$P$, has the same functional dependencies as ours. In the regime of
low $f_{mol}$ , where
Elmegreen's treatment is meant to be appropriate, $f_{mol} \simeq
R_{mol}$.   Although $j$ probably does not have a simple power law
dependence on $\Sigma_{gas}$, if it can be approximated as $j \propto
{\Sigma_{g}}$, then Elmegreen's (1993) expression for $f_{mol}$ has
a power law dependence quite close to what we find observationally.

We note further that although one expects $R_{mol}$ to have a
dependence on $j$ on physical grounds, the small scatter of the lines
in Figure \ref{pressplot} implies that {\it empirically}, $R_{mol}$
behaves as if it depends on pressure only. The discussion above
suggests that the dependence on $j$ is subsumed within the empirical
relationship.

\subsection{Notes on Individual Galaxies}

{\it The Milky Way} -- For the Galaxy, we use the values of
\citet{sigma_dame} for $\Sigma_{\mathrm{HI}}$ and
$\Sigma_{\mathrm{H2}}$ at $R_{gal}>3$~kpc.  We assume that the local
stellar surface density in the vicinity of the Sun is 35
$M_{\odot}$~pc$^{-2}$ and that the stars are distributed in an
exponential disk with scale length 3 kpc \citep{db98} with
$R_0=8.0$~kpc.  We scale $\Sigma_{\mathrm{H2}}$ by 1.36 to account for
the presence of helium and by (2/2.3) to account for the different
conversion factor adopted in that study.  Finally, we scale
$\Sigma_{\mathrm{HI}}$ by 1.5 to account for the presence of high
latitude gas, based on the results beyond the Solar circle from
\citet{levine-h1}.

{\it NGC 598 (M33)} -- NGC 598 is the closest extragalactic source in
our sample.  The disk of NGC 598 is stable according to the Toomre
criterion \citep{mk01}, making it a notable test case for for
comparing the role of gravitational instability with that of pressure
in regulating the ISM and star formation.  We generated a molecular
gas map from a masked integrated-intensity map from the FCRAO 14-m
survey of the galaxy
\citep{m33-fcrao}.  Apart from the Milky Way, it is the only other
disk galaxy in the sample with sufficient surface brightness
sensitivity in the CO map to measure $R_{mol}$ where $R_{mol}<1$.  The
derived index of $\alpha=0.87$ can be compared to the work of
\citet{m33-fcrao} who show that $R_{mol}\propto P_{ext}^{2.3}j^{-1}$
where $j$ is the mean radiation field.  Using $j\propto \Sigma_\star$,
we reproduce their index to the pressure relationship.  Performing a
similar analysis for the remainder of our sample finds indices ranging
from 1.4 to 2.3.
%{\it NGC 3521} --

{\it NGC 3627} -- This galaxy is a molecule rich member of the Leo
Triplet and the ISM of the galaxy has been affected by interaction
with the remainder of the group.  NGC 3628, another member of the
triplet, is also very molecule rich \citep{young-fcrao}.  The
interactions between the three galaxies have stripped a long plume of
material from NGC 3628 and have also likely depleted the \ion{H}{1} in
the disk of NGC 3627.

{\it NGC 4321 (M100)} -- This galaxy is a member of the Virgo cluster,
but is not obviously undergoing ram pressure stripping at the present
time.  Azimuthally averaging the \ion{H}{1} data shows a slight
downturn at large radii, and the median surface density for the
\ion{H}{1} is relatively high (6.7 $M_{\odot}~\mbox{pc}^{-2}$)
compared to the other interacting galaxies in our sample (NGC 3627 and
NGC 4501). Of the three galaxies undergoing interactions, this galaxy
appears to be the least affected by interaction.  However, it is
noticeably deficient in \ion{H}{1} with a total atomic gas mass a
factor of three lower than comparable field galaxies \citep{kenney89}.

%{\it NGC 4414} --

{\it NGC 4501 (M88)} -- We used archival BIMA data from NGC 4501 to
generate the CO map of NGC 4501 used in this study.  The data were
originally presented in \citet{wb02}.  NGC 4501 is a member of the
Virgo Cluster and its \ion{H}{1} morphology appears to be noticeably
affected by the cluster environment.  It is separated by $\sim 500$
kpc on the plane of the sky from M87; and the \ion{H}{1} surface
density is asymmetrical, apparently piled up on the side of the galaxy
closest to M87 \citep{virgo-vla}.  The cluster environment appears to
be responsible for the depletion of the atomic gas which is a factor
of 3 lower than what would be expected from a field spiral of similar
stellar mass \citep{kenney89}.

{\it NGC 4736 (M94)} -- This galaxy has the earliest morphological
type in our sample and sizable amounts of neutral gas are found within
the bulge of the galaxy.  The pressure formulae used in this paper is
appropriate for the disks of galaxies, but should be an upper limit
for the bulges of galaxies, since the thickness of bulges is larger
than disks.  Applying the pressure formulae for disks to NGC 4736 may
account for the high value of $P_0$ in the system.

%{\it NGC 5033} --

% {\it NGC 5055 (M63)} --

{\it NGC 5194 (M51a)} -- NGC 5194 is a classic example of grand-design
spiral structure in galaxies.  The high resolution observations of NGC
5194 facilitate the measurement of $P_{ext}$ and $R_{mol}$ both in and
out of the spiral arms seen in the ISM.  In both the arms and the
interarm regions, $P_{ext}$ appears to be a good predictor of
$R_{mol}$.

{\it NGC 5457 (M101)} -- The overlap between \ion{H}{1} and CO
detections spans only a very small portion of the galaxy.  The
\ion{H}{1} surface density drops markedly in the center of the
galaxy where the molecular gas is concentrated \citep{wb02}.  In the
small overlap region, however, the data from NGC 5457 agree well with
the remainder of sample.

%{\it NGC 7331} --
{\it IC 10} -- We include IC 10 in our sample to investigate the
$R_{mol}$ vs. $P_{ext}$ relationship in a low metallicity environment.
The gas geometry of the system is quite complex and not apparently
confined to the stellar disk, which may affect the applicability of
Equation \ref{pext}.  However, \citet{leroy-ic10} noted that the
molecular gas in the system tended to be found on filaments of atomic
gas that were coincident with the stellar disk of the galaxy.  We
include the pressure data from their work in this study.  They assume
a characteristic scale height of the stellar component of 300 pc.  We
correct the surface densities of gas by the inclination of the stellar
disk assuming that the gas spatially coincident with the stellar disk
actually lies within the disk.

\subsection{Systematic Effects}

In this section, we discuss how variations in our assumptions and
interpretations affect the results.

Since $\Sigma_{\mathrm{H2}}$ appears in the expressions for both
$R_{mol}$ and $P_{ext}$, it is possible that some of the observed
correlation in Figure \ref{pressplot} arises simply because we are
correlating CO emission with itself.  As we can see from
Figure~\ref{alldata}, however, most of our individual data points  
come from the
low pressure regime ($P < P_0$), where no such correlation is expected.
Thus, this concern is most important
where $R_{mol}\gg 1$.  In that case, the Equation \ref{gequation}, as
plotted in Figure
\ref{pressplot} reduces to:

\begin{equation}
{\rm log} \Sigma_{\mathrm{H2}} - {\rm log}\Sigma_{\mathrm{HI}} =
\alpha~{\rm log}\Sigma_{\mathrm{H2}} + \frac{\alpha}{2}~{\rm log}
\Sigma_\star + const.
\label{approxeq}
\end{equation}

\noindent where $\alpha$ is the quantity we are trying to determine
(assuming that $R_{mol}$ is proportional to $P^{\alpha}$). Since
$\Sigma_{\mathrm{HI}}$ shows little variation compared to either
$\Sigma_{\mathrm{H2}}$ or $\Sigma_\star$, and because log
$\Sigma_{\mathrm{H2}}$ dominates the left hand side of Equation
\ref{approxeq} for large $R_{mol}$, we can absorb  log
$\Sigma_{\mathrm{HI}}$ into the constant.  Because log
$\Sigma_{\mathrm{H2}}$ appears on both sides of Equation
\ref{approxeq}, at least some of the correlation at $R_{mol}\gtrsim 5$
is due to correlating log $\Sigma_{\mathrm{H2}}$ with
itself. The slope of the $P - R_{mol}$ relationship is given
by:

\begin{equation}
\alpha = \frac{\log \Sigma_{\mathrm{H2}}+const.}{\log \Sigma_{\mathrm
{H2}}+
(\log \Sigma_\star)/2}
\label{slope}
\end{equation}

\noindent
As long as the range of values of $(\log \Sigma_\star)/2$ are
comparable to the range of values of
$\log \Sigma_{\mathrm{H2}}$, as is the case for most of the data
presented in Figure \ref{pressplot}, both terms in the denominator
will determine the slope, and thus the value of $\alpha$.  Thus, even
for large values of $R_{mol}$, where some of the correlation in the $P
- R_{mol}$ relation is produced by $\Sigma_{\mathrm{H2}}$ appearing on
both axes, the data still provide information about the functional
form of Equation \ref{gequation}.

To estimate the amount of correlation that arises purely from the
mathematical construction of our analysis, we recalculate $R_{mol}$
and $P_{ext}$ after scrambling the original measurements of
$\Sigma_{\mathrm{HI}},\Sigma_{\mathrm{H2}}$ and $\Sigma_\star$.  This
randomization retains the original data but the surface density
measurements for each disk component no longer correspond to the
values of the other components.  Using the scrambled data for
$R_{mol}$ and $P_{ext}$ we do indeed find a significant correlation
in the randomized data, particularly evident at $R_{mol}\gtrsim 2$.  The
power-law index for the scrambled data is $\alpha=1.5\pm 0.02$,
significantly greater than the derived relationship for the real data.
We conclude that we are observing a physically significant relationship
although we recognize that the data for $R_{mol}\gtrsim 2$ would not
be compelling on their own.  When used in conjunction with the data
for low values of $R_{mol}$, however, the molecule-rich regions of
galaxies extend the trend into high-pressure regions.

Another concern is that variations in the CO-to-H$_2$ conversion
factor may substantially alter our results.  However, the only regions
included in this sample that show good evidence for variations in the
CO-to-H$_2$ conversion factor that are also included in this sample
are the molecule rich regions in the central $\sim 200$ pc of galaxies
\citep[e.g.][]{meier-6946}.  In these regions, $R_{mol}\gg 2$
and the observed correlation is largely the product of correlating CO
emission with itself.  As a result, changes in the CO-to-H$_2$
conversion factor will alter $R_{mol}$ and $P_{ext}$ by roughly the
same amount and the data will likely remain consistent with the
observed trend.  Thus, variations in the conversion factor are
unlikely to substantially affect our results.

The projected resolution at the distance of the galaxies varies by
nearly an order of magnitude across the sample.  There is no
significant correlation between physical resolution and the index of
the derived relationship or the degree of scatter around the mean
trend. To assess the effect of the physical resolution on our results,
we repeat the analysis, convolving the data to a constant physical
resolution of 1.8 kpc.  This degrades the resolution to a constant
scale for all galaxies except for NGC 5033.  We find no significant
change in the results as a result of smoothing to this resolution.  We
conclude that resolution effects do not affect the results
significantly.

The assumption of a constant velocity dispersion is well-justified
\citep{vdks,svdk,vdks2,vanzee-1232}, but there are $\sim 30\%$
variations in velocity dispersion within and among galaxies.  These
differences may account for the offsets and scatter seen in Figure
\ref{pressplot}.

% XXX Still need a discussion of the influence of self-gravitation
% clouds on the pressure formulation.

\section{Discussion}

In nearby galaxies where the dominant state of the gas is atomic, the
atomic gas occurs primarily in filaments
\citep{kim-lmc,eprb03,leroy-ic10}, and the formation of the filaments
appears to precede the formation of the molecular gas
\citep{psp5}.  From \S3 and BR04, we have shown that
hydrostatic pressure alone is sufficient to determine what fraction of
the neutral gas at any location in a galaxy is molecular, and we have
determined the quantitative relationship between pressure and
molecular gas fraction.  We have not been able to show, however, how
the molecular gas collects itself into GMCs, or even if the atomic gas
first collects itself into GMC-sized objects and then becomes
molecular.  This detail may not be important for determining the star
formation rates in galaxies on kiloparsec scales, since the rate at
which stars form from the molecular gas appears to be well determined
on scales larger than a GMC \citep{wb02}, but is relevant to issues
related to the stability of the ISM.

In all of the galaxies we have examined in this paper, pressure is
measured on a scale large compared to the size of a GMC, because of
either the relatively poor resolution of the \ion{H}{1} observations, or
because the galaxies are relatively distant.  We expect that
Equation \ref{finalpress} would break down on a scale comparable to
that of a typical GMC because the pressure within a GMC is
enhanced by its self-gravity and is no longer in pressure equilibrium
with its environment.  We would guess that the scale on which
the relation is valid is about equal to twice a pressure scale height
($\sim$ 400 pc in the solar vicinity of the Milky Way).  The pressure
will be in equilibrium on such a vertical scale, on average, and is
unlikely to vary by very much on a similar scale in the galactic
plane.

\subsection {Applicability to Other Kinds of Galaxies}

Our sample of galaxies is limited to normal spirals and some dwarf
galaxies.  It contains weak Seyferts and galaxies with low level
starbursts and galaxies with metallicities that range from solar
values to values as low as 1/6 solar (IC 10).  It does not include
elliptical galaxies, and S0s, a number of which have been shown to be
molecule rich \citep{young2,young-fcrao}, nor does it include very low
metallicity dwarfs such as I Zw 18, luminous starbursts, strong
Seyferts or merging systems.  Equation \ref{finalpress} may not be
applicable in these systems, but the pressure in the ISM can be
estimated from other means.  Equation \ref{finalpress} uses an
equilibrium approximation of disk structure to estimate $\rho
{v_g}^2$, a quantity that can be determined equally well using a
variety of methods.  In elliptical galaxies, the pressure may be
measured directly from density and temperature information inferred
from x-ray data.  In the bulges of disk galaxies, other constraints
from stellar velocity dispersions and scale heights can be used to
estimate a more accurate stellar mass distribution.  Programs are
underway to to probe the relationship between molecular gas fraction
and ISM pressure in an array of such systems.  The Large and Small
Magellanic Clouds will also be good testbeds for determining the
general applicability of the pressure-molecular fraction relation.  In
these two systems with irregular morphologies, we can measure the
relationship between molecular gas fraction and pressure at high
resolution.

Galaxies with low metallicities are interesting regions in which to
study the molecular fraction of the ISM.  In these systems, the flux
of dissociating UV photons tends to be higher than in the disk
galaxies that we studied in this sample.  In conjunction with the
volume density of hydrogen, the UV flux establishes the equilibrium
molecular fraction in the ISM of the neutral gas.  Hence, we suspect
systems with low metallicity and high UV flux (such as the SMC)
may be significantly
offset from the $R_{mol}$ -- $P_{ext}$ relationship seen for disk
galaxies with approximately solar metallicity.

\subsection{Molecular Gas and Star Formation}

\citet[][hereafter K98]{k98} has argued that the surface
density of the rate of star formation $\Sigma_{\mathrm{SFR}}$ is
proportional to the sum of the atomic and molecular gas surface
density, $\Sigma_g$ over a wide range of galaxy luminosity.  This
relationship is puzzling because stars form only from molecular
gas.  K98 cites the relative tightness of his plot of
$\Sigma_{\mathrm{SFR}}$ vs. $\Sigma_g$ compared to his plot of
$\Sigma_{\mathrm{SFR}}$ vs. $\Sigma_{\mathrm{H2}}$. However, when
averaged over kiloparsec scales, $\Sigma_{\mathrm{HI}}$ tends to
saturate at a value of $\sim$10 $M_\sun$ pc$^{-2}$, or about $1.3\times
10^{21}$~cm$^{-2}$ \citep[e.g.~][]{cayatte94,wb02}, the relationship  
between
$\Sigma_{\mathrm{SFR}}$ and $\Sigma_{\mathrm{HI}}$ in K98 is less a
correlation than a plot showing the saturation of the
\ion{H}{1} line.

\citet[][hereafter WB02]{wb02}, on the other hand, have shown
that for 7 well-studied galaxies in the BIMA SONG CO survey,
\citep{song2}, the proportionality found by K98 does indeed hold, but
for
the molecular gas only. That is, K98 finds that $\Sigma_{\mathrm
{SFR}} = 0.16\
\Sigma_g^{1.4}$, where $\Sigma_{\mathrm{SFR}}$ is measured in
$M_\sun$~pc$^{-2}$~Gyr$^{-1}$ and ${\Sigma_g}$ is measured in
$M_\sun$~pc$^{-2}$.  WB02 find that $\Sigma_{\mathrm{SFR}} = 0.13\
\Sigma_{\mathrm{H2}}^{1.4}$.  These relations use values of  $X_
{\mathrm{CO}}$
that differ in whether or not they include a correction for helium.
If helium is included in the WB02 relation, then their coefficient is
0.18 rather than 0.13, which differs from the K98 value by only
13\%.  This is not entirely surprising, since WB02 use K98's
calibration of \ha~line strength vs.~star formation rate.

\citet{mur02} developed a star formation formulation
based on radio continuum measurements calibrated to the supernova rate
in a number of galaxies \citep{condon_araa}, This relation, like WB02,
relates the star formation rate to $\Sigma_{\mathrm{H2}}$ and finds a
power-law index of 1.3, indistinguishable from 1.4 within the
uncertainties.  The \citet{mur02} coefficient of proportionality is a
bit higher: 0.26 rather than 0.18, but uses the \citet{miller-scalo}
IMF rather than a \citet{salpeter} IMF between 0.1 and 100 $M_\sun$.
\citet{mur02} do not give a lower mass limit, but the differences
between the two IMFs could make up much of the discrepancy in the
coefficient.

The K98 relation is based primarily on global averages, and the
relation of WB02 is based on pixel-by-pixel measurements convolved to
the same linear resolution.  As a practical matter, the differences
between these formulations are unimportant for regions of galaxies
where the ISM is dominated by molecular gas, such as galactic centers,
as well
as most of the regions of galaxies over which BIMA SONG has enough
sensitivity to detect CO.  However, throughout the entire disk of the
Milky Way, $\Sigma_{\mathrm{HI}} \ge \Sigma_{\mathrm{H2}}$
\citep[except for the central 300 pc;][]{sigma_dame}, as is also the
case in M33 \citep{m33-fcrao} and most of M31 \citep{iram-m31-aa}.  In
general, the interstellar medium in dwarf galaxies is dominated by
\ion{H}{1} rather than molecular gas, and it is in M33 and in dwarf
galaxies that \citet{mk01} find that their star formation prescription
breaks down.  Observations of \ion{H}{1} rich dwarf galaxies should be
a good test of whether the K98 or WB02 relation is better at
predicting the star formation rate.  One recent study has been
completed by \citet{m33-fcrao} on the Kennicutt-Schmidt law in M33
where they find $\Sigma_{\mathrm{SFR}}\propto
\Sigma_{\mathrm{H2}}^{1.36}$ in good agreement with the work of WB02
and \citet{mur02}.  When they include atomic gas in their analysis,
they find $\Sigma_{\mathrm{SFR}}\propto \Sigma_{\mathrm{H2}}^{3.3}$,
differing dramatically from the results of K98.

\subsubsection {CO, HCN, FIR and Star Formation}
\label{COHCN}
In her ground breaking work on the relationship between dense gas and
star formation, \citet{lada-dense} showed that in the Orion B
molecular cloud (L1630), nearly all of the star formation is
associated with the H$_2$ traced by the CS molecule, a tracer of dense
molecular gas.  Most of the molecular gas in L1630 traced by CO, a
moderate density tracer, is inert and does not take part in the star
formation process.  Why, then, is there a good relationship between
$\Sigma_{\mathrm{CO}}$ and star formation if
the molecular gas traced by CO has little
direct bearing on star formation?

The work of \citet[][hereafter GS04]{gs04} and of \citet{wu05-hcn}
suggests an answer.  GS04 show that HCN is a
linear tracer of the star formation rate in luminous and ultraluminous
galaxies while CO is not. In their work, the star formation rate is
determined from the far IR luminosity.  The connection to Lada's
\citeyearpar{lada-dense} work is that both HCN and CS are good tracers
of dense gas with $n(\mbox{\h2}) \sim 10^{4-5}$, but HCN is somewhat
more luminous, on average, in galaxies \citep[][and references
therein]{hcn-helfer}.  The GS04 plots of the correlation between HCN
and CO with IR luminosity show what appear to be comparably good fits
to the data, though the fit using CO is not linear, while the fit
using HCN is. The correlation coefficients for both fits are
quite high, but the fit using HCN is marginally better than for CO
($R=0.94$ vs. 0.88), and the scatter somewhat less (0.25 dex vs.~0.35
dex).  In luminous and ultra-luminous infrared galaxies, GS04 find
that the HCN luminosity is related to the CO luminosity:
$L_{\mathrm{HCN}} = 1.6 \times 10^{-5}~L_{\mathrm{CO}}^{1.38}$ for $L$
measured in K~km~s$^{-1}$~pc$^2$.

\citet{wu05-hcn} show that the linearity between the HCN and
IR luminosity continues down to the scale of individual dense cores in
Galactic molecular clouds, implying that the HCN -- IR relationship for
starbursts continues down to the level of star-forming clumps. The
linearity holds over a total range of luminosity of 7 -- 8 orders of
magnitude.  Correlations using luminosity usually look remarkably good
because one is correlating distance (squared) with itself on both
axes, making the correlation much more striking than it actually is.
A better comparison is between distance independent quantities such as
surface brightness.  Nevertheless, there seems to be good evidence
that the linearity of IR luminosity with HCN line strength is quite
good for a range of objects from individual Galactic GMCs to ULIRGs.

\subsubsection{A Pressure-Based Prescription for Star Formation on
Galactic Scales}
\label{sfprescription}

We propose an alternative, pressure-based
prescription for galaxy-scale star
formation centering on three empirically-derived relationships. First,
we assume that the FIR luminosity is a good tracer of the massive star
formation in a galaxy because the FIR emission is proportional to the
number of Lyman continuum photons, nearly all of which are absorbed by
dust, From GS04 (and references therein),
\begin{eqnarray}
\dot{M}_{\mathrm{SFR}} & \approx &
2 \times 10^{-10}~(L_{\mathrm{FIR}}/L_{\sun})~~
M_{\sun}~\mbox{yr}^{-1},~\mbox{and} \\
\label{sfrfir}
M_{\mathrm{dense}} &=& 14~\left(\frac{L_{\mathrm{HCN}}}
{\mbox{K km s}^{-1}\mbox{ pc}^{2}}\right)~~M_{\sun}
\label{mdense}
\end{eqnarray}
Next, we assume that all star
formation takes place in dense molecular gas which has a constant star
formation efficiency.
Expressing this in a form comparable to that of
K98 using the above notation gives:
\begin{equation}
\Sigma_{\mathrm{SFR}} = \epsilon \Sigma_{\mathrm{H2}}(\mathrm{HCN})
\label{dense-eff}
\end{equation}
GS04 and \citet{wu05-hcn} measure $\epsilon\approx 10 - 13 \mbox{
Gyr}^{-1}$.
Expressing Equation~\ref{dense-eff} in terms of the total molecular gas
traced by CO gives:
\begin{equation}
\Sigma_{\mathrm{SFR}} = \epsilon \Sigma_{\mathrm{H2}}({\mathrm{CO}})
\left[\frac{\Sigma_{\mathrm{H2}}(\mathrm{HCN})}
{\Sigma_{\mathrm{H2}}(\mathrm{CO})}\right]
\end{equation}
which is related to the total gas content:
\begin{equation}
\Sigma_{\mathrm{SFR}} = \epsilon \Sigma_g f_{mol}
\left[\frac{\Sigma_{\mathrm{H2}}(\mathrm{HCN})}
{\Sigma_{\mathrm{H2}}(\mathrm{CO})}\right]
\label{sfr-ratio}
\end{equation}
where $f_{mol} = \Sigma_{\mathrm{H2}}/{\Sigma_g} = R_{mol}/(R_{mol} +
1)$
and $R_{mol} = \Sigma_{\mathrm{H2}}/\Sigma_{\mathrm{HI}}.$ Third, we
assume that $f_
{mol}$
is determined from the pressure and the total amount of gas present, and
can be calculated from Equations~~\ref{generic-fcn}~and \ref{allgals} .
We can rewrite Equation~\ref{generic-fcn}:
\begin{equation}
f_{mol} = \frac {R_{mol}}{(R_{mol} + 1)} =
\left[1 + \left(\frac {P_{ext}}{P_0}\right)^{-\alpha}\right]^{-1}.
\label{fmol}
\end{equation}

For an integrated intensity, $I$, expressed in units of K~km~s$^{-1}$
and $\Sigma$ measured in $M_\sun~\mbox{pc}^{-2}$,
$\Sigma_{\mathrm{H2}}(\mathrm{HCN}) = 14~I(\mathrm{HCN})$ (GS04), and
$\Sigma_{\mathrm{H2}}(\mathrm{CO}) = 4.4~I(\mathrm{CO})$ for our
adopted CO-to-H$_2$ conversion factor (\S\ref{derivations}) corrected
for helium.  Thus,
\begin{equation}
\frac{\Sigma_{\mathrm{H2}}(\mathrm{HCN})}
{\Sigma_{\mathrm{H2}}(\mathrm{CO})}  =  3.2
\frac{I(\mathrm{HCN})}{I(\mathrm{CO})}
\label{hcn-co}
\end{equation}

Unlike theoretical explanations of the Kennicutt-Schmidt law
\citep[e.g.~][]{krumholz05}, this formulation separates the
empirical star formation relationship of K98 into three
constituent empirical relationships.  The ability to reproduce the K98
results should not be surprising -- all three of the constituent
relationships are based on galaxies similar to those in K98.  A
consistent constant of proportionality is an indication of accurate
calibration of the various quantities involved.  What is notable about
this formulation is the ability to characterize star formation on a
galactic scale using the star formation efficiency of dense gas and
the role of pressure in determining both the  fraction of
gas that is molecular and the fraction of that molecular gas that is
dense enough to form stars.

%\subsubsubsection {The Low Pressure Regime}
\centerline{\it The Low Pressure Regime}

Except for the central 300 pc of the Milky Way, and the inner regions
of M31, the mean \ion{H}{1} surface density exceeds that of the H$_2$  
when
averaged over several hundred pc everywhere in each galaxy.  Thus,
most of the molecular gas in these galaxies occurs in the regime where
$P_0 > P_{ext}$.  For most GMCs in Local Group galaxies, the
surface density of GMCs as traced by CO is roughly constant with a
value of $\sim 100~M_\sun~\mbox{pc}^{-2}$
\citep{psp5}, so the fraction of dense gas and consequently, the star
formation rate of individual GMCs, will also be roughly
constant, as is observed \citep{ms88,sg89}.  Variations in
$\Sigma_{\mathrm{SFR}}$ over kiloparsec scales thus result from
changes in the fraction of neutral gas bound up in GMCs ($\propto
f_{mol}$).  That is, averaging emission on large enough scales
($\gtrsim$ 200 pc) over which the surface filling fraction of molecular
gas is small, pushes the Kennicutt-Schmidt law into the regime of small
values of $\Sigma_{\mathrm{H2}}$ since averaging GMCs with atomic gas
that has no star formation will geometrically dilute
both $\Sigma_{\mathrm{SFR}}$ and $\Sigma_{\mathrm{H2}}(\mathrm{CO})$
by a comparable, though not identical, factor.

In this regime, we may take Equation~\ref{hcn-co} to be constant.
\citet{hcn-MW} derive a value of 0.026 $\pm 0.008$ for the ratio of
integrated intensities averaged over the inner disk of the Milky Way.
Similarly, GS04 find $L(\mathrm{HCN})/L(\mathrm{CO})=0.039$ for their
low-luminosity galaxies.  We adopt the average of these two values as
a characteristic line ratio which implies that the ratio in
Equation \ref{hcn-co} is 0.1.  Using this and taking $P_{ext} \ll
P_0$, Equation~\ref{sfr-ratio} becomes
\begin{equation}
\Sigma_{\mathrm{SFR}} = 0.1 ~\epsilon \Sigma_g
\left(\frac{P_{ext}}
{P_0}\right)^{0.92}~~M_\sun~\mbox{pc}^{-2}~\mbox{Gyr}^{-1}
\label{sfr-lowpress}
\end{equation}
where $P_{ext}$ is given by Equation~\ref{pext} and P$_0/k = 4.3
\times 10^{4}$ cm$^{-3}$ K.
This expression will, in general, give smaller values of
$\Sigma_{\mathrm{SFR}}$ than K98.

%\subsubsubsection{The High Pressure Regime}
\centerline{\it The High Pressure Regime}

For galaxies where $R_{mol} \gg 1$ over large scales in the ISM, the
situation changes. In this case, $P_{ext} > P_0$, and
$f_{mol} \approx$~1. In at least part of this regime, GMCs do not have
a constant surface density, and one cannot take Equation \ref{hcn-co}
to be constant.  For $f_{mol} \gtrsim~0.8, \Sigma_g \approx
\Sigma_{\mathrm{H2}}(\mathrm{CO})$.  Then the observed Schmidt-law
relationship in the molecular gas \citep[$\Sigma_{\mathrm{SFR}}
\propto {\Sigma_{\mathrm{H2}}}^{1.4}$;][]{wb02,mur02,m33-fcrao} reduces
to the prescription of K98, $\Sigma_{\mathrm{SFR}} =
0.16~{\Sigma_g}^{1.4}$.

An alternate expression can be derived from the consideration that
stars form only from dense gas.  GS04 and \citet{wu05-hcn} propose a
star formation prescription based on the dense gas content of a
region:
\begin{equation}
\dot{M}_{\star} = (1.1\pm 0.2) \times 10^{-8}
M_{\mathrm{dense}}~M_{\odot}~\mbox{yr}^{-1}
\end{equation}
where $M_{dense}$ is the mass of dense gas including helium measured
in solar masses.  Using the scaling between $L_{\mathrm{HCN}}$ and
$L_{\mathrm{CO}}$ used in the GS04 study, this suggests the star
formation rate can be deduced from the total molecular gas mass of the
galaxy:
\begin{equation}
\dot{M}_{\star} = (0.77 \pm 0.07)
\left(\frac{M_{\mathrm{H2}}}{10^9~M_{\odot}}\right)^{1.44}~
M_{\odot}~\mbox{yr}^{-1}.
\label{sfr-hipress}
\end{equation}
In the high-pressure regime, $M_{\mathrm{H2}}\approx M_{g}$ so
$\dot{M}_{\star}\propto M_{g}^{1.4}$.  While the index of 1.4 is
reminiscent of that found in K98, the two expressions are
qualitatively different. Equation \ref{sfr-hipress} deals with the
global properties of a system (mass) rather than the local properties
(surface density) used in K98.

To what degree are the prescriptions, Equations~\ref{sfr-lowpress} and
\ref{sfr-hipress}, both locally and globally applicable?  Both
equations use relations that are globally determined
(Equations~\ref{sfrfir} and \ref{dense-eff}), so they will both be
applicable globally.  But both equations also use locally determined  
relations
(Equations~\ref{allgals} and \ref{sfr-ratio}).  As long as care is
taken to normalize the global relations over the same area, the basic
relations should be applicable locally as well (over $\sim$ kpc scales).
Equation~{\ref{sfr-hipress} is determined from CO observations that
include luminous IR galaxies (LIRGS - GS04), so should be applicable  
to these
systems as well.  However, because a scaling has not been established
for ULIRGS, care should be taken when applying our relations to these
systems.  We use these relations both locally and globally in Section
\ref{comps} below.

%A star formation prescription that depends on the local properties of
%equires detailed knowledge of the behavior of
%$I(\mbox{HCN})/I(\mbox{CO})$ in extreme physical conditions.  The work
%of \citet{hcn-MW} and GS04 suggest that the ratio is roughly constant
%in the low pressure regime.  However, the non-linear scaling between
%$L(\mbox{HCN})$ and $L(\mbox{CO})$ found by GS04 at large values of
%$L(\mbox{CO})$ implies that this ratio is not constant in all
%environments.

\subsubsection{Comparison with Observations}
\label{comps}

In this section, we compare the results of our star formation
prescription to observed estimates of the star formation rate in
various galaxies.

{\it The Milky Way} -- We estimate the star formation rate in the
Milky Way using Equation \ref{sfr-ratio}.  We express $f_{mol}$ in
terms of the pressure (Equation \ref{fmol}); because the disk of the
Galaxy is entirely in the low pressure regime, we assume a constant
dense gas fraction: $\Sigma_{\mathrm{H2}}(\mathrm{HCN})/
\Sigma_{\mathrm{H2}}(\mathrm{CO})=0.1$
\citep[GS04]{hcn-MW}.  This gives an expression for $\Sigma_{\mathrm 
{SFR}}$
as a function of Galactic radius.  We plot the prediction in
Figure \ref{mw-sfr-prediction}.  The surface density of star formation
expected from the K98 prescription is shown as a dashed curve,
$\Sigma_{\mathrm{SFR}} =0.16~{\Sigma_g}^{1.4}~
M_\sun~\mbox{pc}^{-2}~\mbox{Gyr}^{-1}$. We also compare the results to
the observed data for \ion{H}{2} regions \citep{guesten82},
pulsars \citep{lyne85}, and supernova remnants \citep{guibert78},
as plotted by \citet{prantzos95},
normalizing to a local star formation rate:
$\Sigma_{\mathrm{SFR}}(R_0) = 1.9~M_{\odot}\mbox{ pc}^{-2}\mbox{
Gyr}^{-1}$ \citep{kroupa95}.
In general, the pressure-based prediction agrees reasonably well with  
the
observed data.  In contrast, the K98 prediction gives good agreement
at the peak of the molecular ring ($R_{gal} \sim$ 4 kpc).
but does not predict the
shape or the amplitude of the curve well at larger $R_{gal}$;
it overpredicts the SFR,
as expected. Neither does particularly well inside the peak of the
molecular ring, but that may be due to uncertain kinmatics induced by
the bar.  In the Solar neighborhood, we
predict $\Sigma_{\mathrm{SFR}}=2.2~
M_\sun~\mbox{pc}^{-2}~\mbox{Gyr}^{-1}$ compared to
1.9 $M_\sun~\mbox{pc}^{-2}~\mbox{Gyr}^{-1}$ inferred from
\citet{kroupa95} and $3-5 ~
M_\sun~\mbox{pc}^{-2}~\mbox{Gyr}^{-1}$ summarized in \citet{rana91}.
Integrating under the curve, the pressure-based prescription
gives a total star formation rate for the Milky Way of $\dot{M}_\star
= 1.1~M_{\odot}~\mbox{yr}^{-1}$ compared to 2.0 for the K98
formulation.  Integrating the observed data over
the same range gives
$\dot{M}_\star = 1.5~M_{\odot}~\mbox{yr}^{-1}$.

\begin{figure}
\begin{center}
\plotone{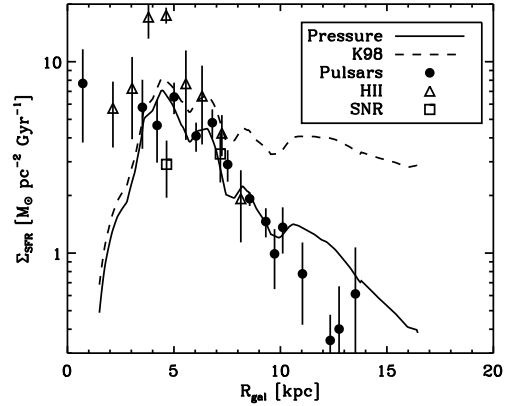}
\caption{\label{mw-sfr-prediction} Star formation rate as a function
of galactocentric radius ($R_{gal}$) for the Milky Way.  The figure
compares the results for the pressure-based star formation
prescription in the present work (solid) to that of K98 (dashed) and
the measured values. Triangles: \ion{H}{2} regions \citep{guesten82};
Circles: Pulsars \citep{lyne85}; Squares: SNR \citep{guibert78}.
The pressure formulation predicts not only the shape of the curve for
the observed star formation rate, but for most of the points,
the amplitude as well.}
\end{center}
\end{figure}

{\it NGC 598 (M33)} -- We perform a similar analysis on M33 using the
star formation rate derived from infrared data by \citet{m33-fcrao}.
Again, M33 is entirely in the low pressure regime.
The integrated value of star formation in the galaxy is 0.7
$M_{\odot}~\mbox{ yr}^{-1}$ in good agreement with the values found
using other methods \citep{m33-iso}.  In Figure \ref{m33-obs}, we
compare the observed profile of star formation to that predicted by
the pressure-based formalism and that of K98. Once again, both the
pressure-based formalism and that of K98 underpredict the amount of
star formation.  As previously noted by \citet{m33-fcrao}, the K98
prescription does a poor job of predicting the slope of the star
formation rate in M33.  The pressure-based prediction agrees with the
slope of the observational data, only the values are low by a factor
of seven.  Yet Table \ref{results} shows that the pressure
formulation predicts the molecular gas fraction reasonably well in
M33.  The discrepancy must arise in the efficiency of star
formation in the dense gas, $\epsilon$, the dense gas fraction,
$\Sigma_{\mathrm{H2}}(\mbox{HCN})/ \Sigma_{\mathrm{H2}}(\mbox{CO})$
or the calibration of either the CO or the H$\alpha$
data.  Future observational work will be required to determine why
the molecular gas in M33 appears to be so efficient at forming stars.

\begin{figure}
\begin{center}
\plotone{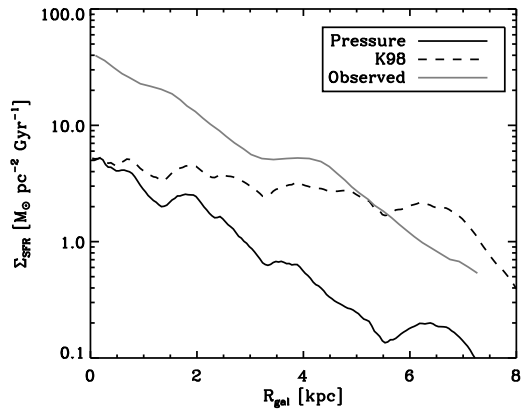}
\caption{\label{m33-obs} Star formation rate as a function
of galactocentric radius ($R_{gal}$) for NGC 598 (M33).  The figure
compares the results for the pressure-based star formation
prescription in the present work (solid) to that of K98 (dashed) and
the measured values \citep[gray,][]{m33-fcrao}.  The pressure
formulation predicts the shape of the observed curve, though the
results are offset by a constant factor of 7, implying the star
formation efficiency of molecular gas in M33 is dramatically higher
than in comparable systems.}
\end{center}
\end{figure}

{\it Molecule-Rich Galaxies} -- Finally, we examine the predictions of
the pressure-based star formation prescription in the molecule-rich
galaxies of \citet{wb02}. In these galaxies, the measured
${\Sigma_{\mathrm{H2}}(\mathrm{CO})}$ is everywhere greater than
${\Sigma_{\mathrm{HI}}}$, so we use Equation~\ref{sfr-hipress} to
determine the SFR.  We compare the K98 prescription and
pressure-based results  for six galaxies in Figure \ref{molrich-pred}.
We omit M101 (NGC 5457) from our analysis because of the narrow range
of $R_{gal}$ over which all relevant quantities are measured.  The
observed star formation data are those of \citet{wb02} for the case
where $A_V\propto N(\mbox{H})$. For these galaxies,
the pressure-based star formation
prediction agrees well with that of K98 since the galaxies are
predominantly molecular.  The two predictions agree reasonably well
with observed data except for NGC 4321.
The \citet{wb02} observations are insufficiently sensitive to go into  
the low
pressure regime where
the pressure-based formulation may do better at reproducing
the observed results.  Indeed, it is when $R_{mol} \lesssim 2$ that
the relations diverge, which will occur for the outer regions of large
spirals, \ion{H}{1}-rich dwarfs such as dwarf irregulars, and damped
Ly-$\alpha$ systems.  For these objects, our predicted star formation
rates are significantly lower than those of \citet{k98}.  Since dwarf
irregulars often show episodic star formation, our prescription
assumes that enough dIrrs are observed to be representative of the
population as a whole.

\begin{figure*}
\begin{center}
\plotone{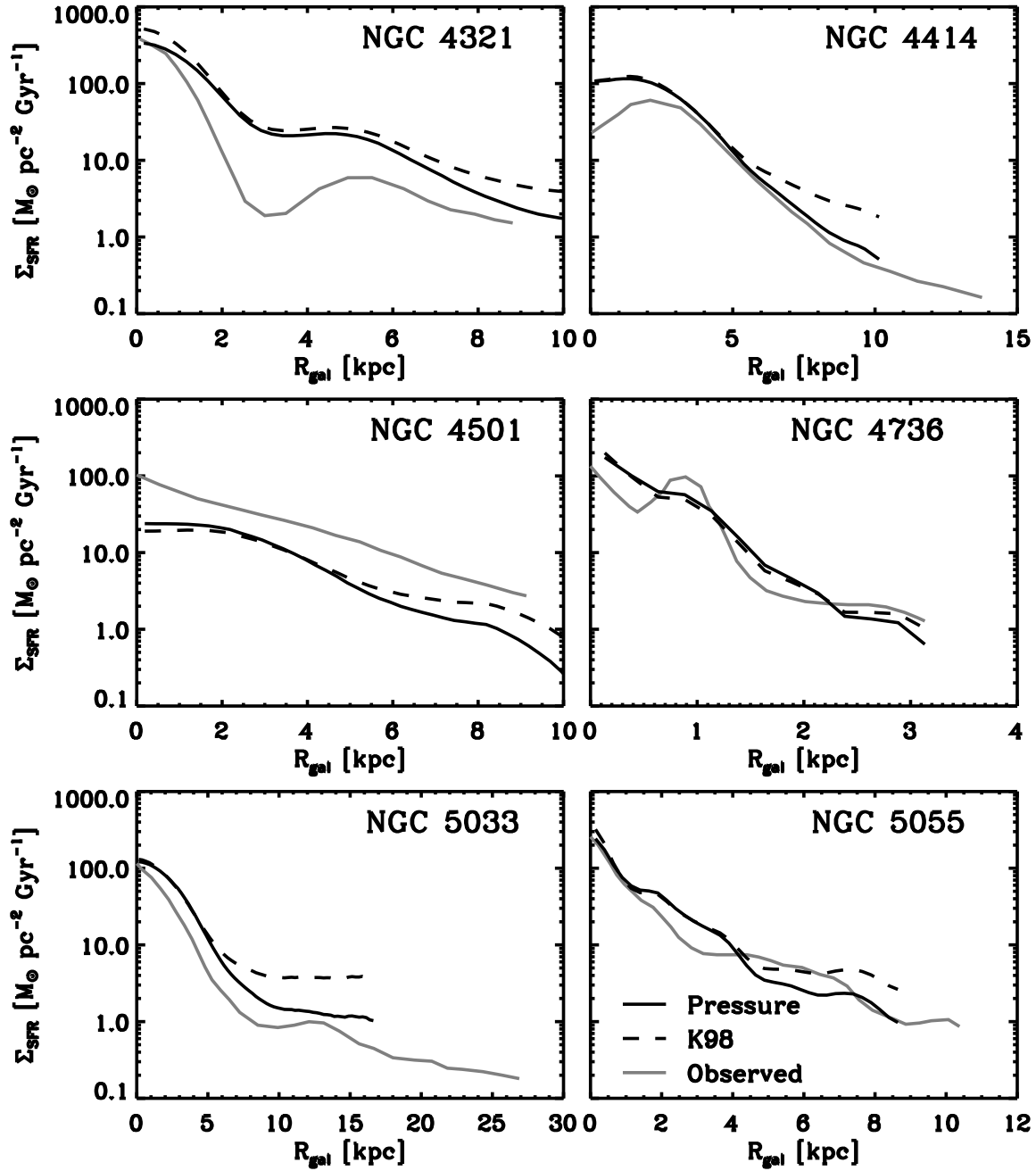}
\caption{\label{molrich-pred} Star formation rate as a function
of galactocentric radius ($R_{gal}$) for six molecule-rich galaxies
studied in \citet{wb02}.  For each panel, the observed star formation
rate is shown as a gray curve, the pressure-based prediction as a
solid curve and the K98 prediction as a dashed curve.  All three
curves agree reasonably well, but the pressure prediction tends to
reproduce results in the outer regions of the galaxies well.}
\end{center}
\end{figure*}

We also compare the observed global star formation rates in these
Table~\ref{wb02table}. The global star formation rates given in the
table are determined by integrating under each of the curves.  On the
basis of the data in this table, there is no significant difference in
using K98 or our Equation~\ref{sfr-hipress} to predict star formation
rates for these large, molecule-rich galaxies, though the
pressure-based prescription tends to give somewhat higher values.  A
difference of about 15\% is expected as a result of a somewhat higher
value of the coefficient in Equation~\ref{sfrfir} compared to
the K98 value.

%\documentclass[preprint]{aastex}
%\begin{document}
\begin{deluxetable}{ccccc}
\tablecaption{Results}
\tablewidth{0pt}
\tablehead{
\colhead{Galaxy} & \colhead{$M_{\mathrm{H2}}$} & \colhead{$\dot{M}_  
{\star}$, Observed} &
\colhead{$\dot{M}_{\star}$, Pressure} &
\colhead{$\dot{M}_{\star}$, K98}\\
\colhead{} & \colhead{($10^9~M_{\odot}$)} &
\colhead{($M_{\odot}~\mbox{yr}^{-1}$)} &
\colhead{($M_{\odot}~\mbox{yr}^{-1}$)} &
\colhead{($M_{\odot}~\mbox{yr}^{-1}$)}
}
\startdata
NGC 4321 & 6.5 & 1.8 & 10.4 & 6.6 \\
NGC 4414 & 4.4 & 3.0 & 6.2 & 4.9 \\
NGC 4501 & 1.6 & 3.2 & 1.4 & 1.2 \\
NGC 4736 & 0.3 & 0.3 & 0.2 & 0.4 \\
NGC 5033 & 4.2 & 1.9 & 5.2 & 3.8 \\
NGC 5055 & 1.6 & 1.2 & 1.5 & 1.7 \\
\enddata
\label{wb02table}
\end{deluxetable}
galaxies
%where the neutral gas is predominantly atomic (the low pressure
%regime), GMCs are very similar
%from galaxy to galaxy and have similar star formation efficiencies.
%The external pressure in the ISM establishes $R_{mol}$ for each galaxy
%accounting for the range of $\Sigma_{\mathrm{H2}}$ reported in K98.
%The Kennicutt-Schmidt law correlation in these systems is the result
%of averaging many GMCs with relatively a constant internal pressure
%and thus constant dense gas fraction and constant star formation rate
%with inert atomic gas over kiloparsec scales.
%In molecule-rich regions, the GMCs are influenced
%by the high-pressure environment and have higher fractions of
%$\Sigma_{\mathrm{H2}}(\mathrm{HCN})$. accounting for the enhanced star
%formation rates.  The relationship between $L_{\mathrm{H2}}{\mathrm
%{HCN}}$
%$ \propto L_{\mathrm{h2}}{\mathrm{CO}}^{1.4}$ accounts for the
%1.4 power law index in the Kennicutt-Schmidt star formation law.

\subsubsection {Comparison with Simulations}

A fundamental difference between the two relations comes when one
wishes to compare the predictions of the star formation rate from
numerical simulations with that of the star formation prescriptions.
In the K98 formulations, since the star formation rate depends only on
the total amount of gas, the state of the gas is unimportant.  Here,
we argue that the state of the gas is of central importance, and can
be determined by Equation \ref{finalpress} from the total amount of
gas and knowledge of the gas pressure.  Thus, high column density,
low-pressure gas should have relatively few molecules and therefore a
low star formation rate.  High column density, high-pressure gas
should have about the same star formation rate in both formulations,
because there will be nearly complete conversion of atomic to
molecular gas.

\section {Summary and Conclusions}

\begin{itemize}
\item We show that the ratio of atomic to molecular gas in galaxies is
determined by hydrostatic pressure and that the relationship between
the two is nearly linear, over three orders of magnitude in pressure.
The rms deviation of individual points from the mean relation is a
0.26 dex, or 80\%. The scatter of the binned data for all of the
galaxies is 0.15 dex, only 38\%.
\item In addition to very similar slopes, the galaxies have only a
small range in zero points, except for three galaxies that are
interacting with their environments.
\item The dispersion of measurements within individual galaxies varies
widely from galaxy to galaxy, with some galaxies showing
extraordinarily small dispersions.
\item Although some of the correlation is driven by the appearance
of $\Sigma_{\mathrm{H2}}$ on both axes at values of high $R_{mol}$,
measurement of the power law index of the pressure is nevertheless
reliable.
\item Although the galaxies in our sample are missing S0, E, strong
starburst and Seyferts, we probe a large range in mass, metallicity
and column density, suggesting that the $P_{ext}-R_{mol}$ relation has
wide applicability.
\item Even though star formation is associated with dense gas traced
by CS and HCN, we show that CO is as good a tracer of star formation in
galaxies as HCN, even in starbursts. This is due to the tight,
non-linear relationship between ${\Sigma_{\mathrm{SFR}}}$ and
${\Sigma_{\mathrm{H2}}(\mathrm{CO})}$ in galaxies.
\item We derive a pressure-based prescription for star formation in
galaxies that is based on the $P-R_{mol}$ relation derived in this
paper.  It differs from K98 for the outer regions
of large spirals, dwarf galaxies,
damped Lyman alpha systems and other molecule-poor galaxies.  For
molecule-rich galaxies, the prescription has a form similar to that of
K98.
\end{itemize}

\acknowledgements
We thank Tony Wong and an anonymous referee for helpful comments on
the manuscript.  In particular, the referee pointed out a
non-trivial error in the original manuscript.
We are grateful to Michelle Thornley who provided us with all
\ion{H}{1} data sets for the BIMA SONG galaxies. Pavel Kroupa
gave us the estimate for the local star formation rate based on
his work. LB's work is supported by NSF grant AST 02-28963. ER's work is
supported by NSF grant AST-0502605.

%\bibliographystyle{apj}
%\bibliography{refs}

\end{document}